# Partial-Diffusion Least Mean-Square Estimation Over Networks Under Noisy Information Exchange

Vahid Vadidpour, Amir Rastegarnia, Azam Khalili
Department of Electrical Engineering
Malayer University
Malayer, Iran, 65719-95863
Email: v.vahidpour.eng@ieee.org

Saeid Sanei
Department of Computing
University of Surrey
Surrey GU2 7XH, UK
Email: s.sanei@surrey.ac.uk

*Abstract*—**Partial diffusion scheme is an effective method for reducing computational load and power consumption in adaptive network implementation. The Information is exchanged among the nodes, usually over noisy links. In this paper, we consider a general version of partial-diffusion least-mean-square (PDLMS) algorithm in the presence of various sources of imperfect information exchanges. Like the established PDLMS, we consider two different schemes to select the entries, sequential and stochastic, for transmission at each iteration. Our objective is to analyze the aggregate effect of these perturbations on general PDLMS strategies. Simulation results demonstrate that considering noisy link assumption adds a new complexity to the related optimization problem and the trade-off between communication cost and estimation performance in comparison to ideal case becomes unbalanced.**

## I. INTRODUCTION

Due to limited electrical power and bandwidth resources for internode communication over a practical wireless sensor networks (WSN) or ad hoc networks, data transmission through radio communication links can become prohibitively expensive for realizing a collaborative task. Generally speaking, although benefits of diffusion strategies achieved by increasing internode communications, they are compromised by the communication cost. As a result, since various nodes can have various numbers of neighbors, they may require disparate hardware or consume power dissimilarity [1]. Therefore, reducing the communication cost while maintaining the benefits of cooperation is of practical importance [1].

There have been several attempts to reduce the communication cost without considerable degradation of the estimation and compromising the cooperation benefits in diffusion algorithms. Among them diffusion least mean-square (LMS), such as reducing the dimension of the estimates [2-4], selecting a subset of the entries of the estimates [5, 6], set-membership filtering [7, 8], or partial updating [9] have been reported in [10-12]. Among these methods, we focus on [5] which LMS algorithm for adaptive distribute estimation has been formulated and analyzed by utilizing partial-diffusion. In [5], an adapt-then-combine (ATC) partial-diffusion least mean-square (PDLMS) algorithm is proposed for distributed estimation over adaptive networks in which, at each iteration, each node transmits a subsets of the entries of intermediate estimate vector to its neighbors.

In the PDLMS strategy proposed in [5], the weight estimates that are exchanged among the nodes can be subject to perturbations over communication links. The effect of link noise during the exchange of weight estimates, already appear for the diffusion algorithm in the works [13-17]. In comparison to the prior work on PDLMS in [5], this manuscript develops a more general class of PDLMS of which [5] is a special case. Like [5], we also consider two different schemes for selecting the entries for transmission at each iteration. It should be noted that since our objective is to minimize the internode communications, the nodes only exchange their intermediate estimates with their neighbors. Therefore, we allow for noisy exchange just during the two combination steps. We subsequently study the performance of this general case utilizing the energy conservation argument [18]. We established its stability and convergence in the mean and mean-square senses. We also derive a theoretical expression for the steady-state mean-square-deviation (MSD) and verify its accuracy via numerical simulations. The analysis further demonstrates that the noises related to the exchange of weight estimates do not change the dynamics of the network but lead to network performance deterioration.

This paper is organized as follows. In Section II, we formulate the PDLMS under noisy information exchange. The performance analyses are examined in Section III. We provide simulation results in Section IV and draw conclusions in Section V.

### A. Notation

We adopt the lowercase letters to denote vectors, uppercase letter for matrices, normal font for nonrandom (deterministic) quantities, and the boldface letters for random quantities. The notation $(.)^*$ refers to conjugate transposition, $Tr(.)$ refer to the trace of its matrix argument, $\otimes$ for the Kronecker product, and $vec(.)$ for a vector formed by stacking the columns of its matrix argument. We shall also use $col(...)$ to denote a column vector formed by stacking its arguments on top of each other and $diag(...)$ to denote a (block) diagonal matrix formed from its argument. All vectors in our treatment are column vectors, with the exception of regression vectors, $\boldsymbol{u}_{k,i}$.

## II. PARTIAL DIFFUSION ALGORITHMS WITH IMPERFECT INFORMATION EXCHANGE

Consider a connected network consisting of $N$ nodes. At time instant $i \geq 0$, each node $k$ has access to scalar measurements $d_k(i)$ and $1 \times M$ regression data vectors $u_{k,i}$. The data across all nodes are assumed to be related to an unknown $M \times 1$ vector $w^o$ via linear regression model of the form [18]:

$$d_k(i) = u_k(i)w^o + v_k(i) \quad (1)$$

where $v_k(i)$ denotes the measurement noise with zero mean and variance $\sigma_{v,k}^2$ and the vector $w^o$ refers to the parameter of interest.

We are now interested in solving optimization problems of the type:

$$\min_w \sum_{k=1}^{N} \mathbb{E}|d_k(i) - u_{k,i}w|^2 \quad (2)$$

The nodes in the network would like to estimate $w^o$ by solving the equation above in adaptive and collaborative manners. We review the diffusion adaptation strategies with imperfect information exchange below.

### A. Diffusion Adaption with Imperfect Information Exchange

Consider the following general adaptive diffusion strategies with $C = I_N$ corresponding to the case in which the nodes only share the weight estimates for $i \geq 0$ [19]:

$$\phi_{k,i} = \sum_{l \in \mathcal{N}_k} a_{1,lk} w_{l,i-1} \quad (3)$$

$$\psi_{k,i} = \phi_{k,i-1} + \mu_k u_{k,i}^*[d_k(i) - u_{k,i}\phi_{k,i-1}] \quad (4)$$

$$w_{k,i} = \sum_{l \in \mathcal{N}_k} a_{2,lk} \psi_{l,i} \quad (5)$$

The scalars $\{a_{1,lk}, a_{2,lk}\}$ are non-negative real coefficients corresponding to the $(l,k)$ entries of $N \times N$ combination matrices $\{A_1, A_2\}$, respectively. The role of these combination matrices is in convergence behavior of the diffusion strategy (3)-(5). These coefficients are zero whenever node $l \notin \mathcal{N}_k$, where $\mathcal{N}_k$ denotes the neighborhood of node $k$. These matrices are assumed to satisfy the conditions:

$$A_1^T \mathbb{1}_N = \mathbb{1}_N, \qquad A_2^T \mathbb{1}_N = \mathbb{1}_N \quad (6)$$

where the notation $\mathbb{1}$ denotes an $N \times 1$ column vector with all its entries equal to one.

We model the noisy data received by node $k$ from its neighbor $l$ as follows (see Fig. 1):

$$w_{lk,i-1} = w_{l,i-1} + v_{lk,i-1}^{(w)} \quad (7)$$

$$\psi_{lk,i} = \psi_{l,i} + v_{lk,i}^{(\psi)} \quad (8)$$

where $v_{lk,i-1}^{(w)}$ ($M \times 1$) and $v_{lk,i}^{(\psi)}$ ($M \times 1$) are vector noise signal. They are temporally white and spatially independent random process with zero mean and covariance given by $\{R_{v,lk}^{(w)}, R_{v,lk}^{(\psi)}\}$. The quantities $\{R_{v,lk}^{(w)}, R_{v,lk}^{(\psi)}\}$ are all zero if $l \notin \mathcal{N}_k$ or when $l = k$. It should be noted that the subscript $lk$ indicates that $l$ is the source and $k$ the sink, the flow of information is from $l$ to $k$.

Using the perturbed data (7) and (8), the adaptive strategy (3)-(5) becomes

$$\phi_{k,i} = \sum_{l \in \mathcal{N}_k} a_{1,lk} w_{lk,i-1} \quad (9)$$

$$\psi_{k,i} = \phi_{k,i-1} + \mu_k u_{k,i}^*[d_k(i) - u_{k,i}\phi_{k,i-1}] \quad (10)$$

$$w_{k,i} = \sum_{l \in \mathcal{N}_k} a_{2,lk} \psi_{lk,i} \quad (11)$$

### B. Partial-Diffusion with Imperfect information Exchange

In order to lower the level of internode communication required among the nodes, we utilize partial-diffusion strategy proposed in [5], to transmit $L$ out of $M$ entries of the intermediate estimates at each time instant where the integer $L$ is fixed and pre-specified. Again, we develop a more general class of PDLMS of which [5] is a special case. The selection of to-be-transmitted entries at node $k$ and time instant $i$ can be portrayed by an $M \times M$ diagonal entry-selection matrix, denoted by $\Lambda_{k,i}$, that has $L$ ones and $M - L$ zeros on its diagonal [5]. The position of ones states the selected entries. Multiplication of an intermediate estimate vector by this matrix replaces its non-selected entries with zero.

According to (9) and (11) that can also be expressed as:

$$\phi_{k,i} = a_{1,kk} w_{k,i-1} + \sum_{l \in \mathcal{N}_k \setminus \{k\}} a_{1,lk}[\Lambda_{l,i-1} w_{lk,i-1} + (I_M - \Lambda_{l,i-1}) w_{lk,i-1}] \quad (12)$$

$$\psi_{k,i} = \phi_{k,i-1} + \mu_k u_{k,i}^*[d_k(i) - u_{k,i}\phi_{k,i-1}] \quad (13)$$

$$w_{k,i} = a_{2,kk}\psi_{k,i} + \sum_{l \in \mathcal{N}_k \setminus \{k\}} a_{2,lk}[\Lambda_{l,i}\psi_{lk,i} + (I_M - \Lambda_{l,i})\psi_{lk,i}] \quad (14)$$

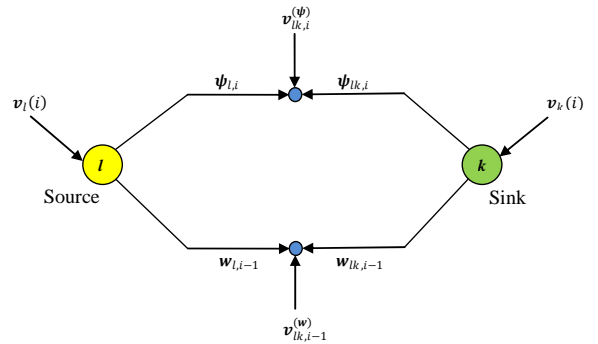

Figure 1. Several additive nise sources perturb the exchange of information from node $l$ to node $k$.

Each node needs the information of all entries of its neighbors' intermediate estimate vectors for the consultation phase. However, when the intermediate estimates are broadcast partially ($0 < L < M$), nodes have no access to the non-communicated entries. To resolve this indistinctness, we allow the nodes utilize their own intermediate estimates entries instead of ones from the neighbors that have not been communicated [5], i.e., at node $k$, substitute

$$(I_M - \Lambda_{l,i-1})w_{k,i-1} \quad \forall l \in \mathcal{N}_k \setminus \{k\} \quad (15)$$

for

$$(I_M - \Lambda_{l,i-1})w_{lk,i-1} \quad \forall l \in \mathcal{N}_k \setminus \{k\} \quad (16)$$

and

$$(I_M - \Lambda_{l,i})\psi_{k,i} \quad \forall l \in \mathcal{N}_k \setminus \{k\} \quad (17)$$

for

$$(I_M - \Lambda_{l,i})\psi_{lk,i} \quad \forall l \in \mathcal{N}_k \setminus \{k\} \quad (18)$$

Based on this approach together with using perturbed data as introduced in (7) and (8), we formulate general PDLMS under noisy exchange as follows:

$$\phi_{k,i} = a_{1,kk}w_{k,i-1}$$
$$+ \sum_{l \in \mathcal{N}_k \setminus \{k\}} a_{1,lk}[\Lambda_{l,i-1}w_{l,i-1} + \quad (19)$$

$$\psi_{k,i} = \phi_{k,i-1} + \mu_k u_{k,i}^*[d_k(i) - u_{k,i}\phi_{k,i-1}] \quad (20)$$

$$w_{k,i} = a_{2,kk}\psi_{k,i}$$
$$+ \sum_{l \in \mathcal{N}_k \setminus \{k\}} a_{2,lk}[\Lambda_{l,i}\psi_{l,i} + (I_M - \Lambda_{l,i})\psi_{k,i}] + \sum_{l \in \mathcal{N}_k \setminus \{k\}} a_{2,lk}\Lambda_{l,i}v_{lk,i}^{(\psi)} \quad (21)$$

*Remark:* The probability of transmission for all the entries at each node is equal and state as [5]

$$p = L/M \quad (22)$$

Moreover, the entry selection matrices, $\Lambda_{k,i}$, do not rely on any data/parameter with the exception of $L$ and $M$.

Introduce the following aggregate $M \times 1$ zero mean noise signals:

$$v_{k,i-1}^{(w)} \triangleq \sum_{l \in \mathcal{N}_k \setminus \{k\}} a_{1,lk}\Lambda_{l,i-1}v_{lk,i-1}^{(w)} \quad (23)$$

$$v_{k,i}^{(\psi)} \triangleq \sum_{l \in \mathcal{N}_k \setminus \{k\}} a_{2,lk}\Lambda_{l,i}v_{lk,i-1}^{(\psi)} \quad (24)$$

These noises correspond to the cumulative effect on node $k$ of all selected exchange noises from the neighbors of node $k$ while exchanging the estimates $\{w_{l,i-1}, \psi_{l,i}\}$ in the course of the two consultation steps. The $M \times M$ covariance matrices of these noises are given by:

$$R_{v,k}^{(w)} \triangleq \sum_{l \in \mathcal{N}_k \setminus \{k\}} a_{1,lk}^2 \Lambda_{l,i-1} R_{v,lk}^{(w)} \quad (25)$$

$$R_{v,k}^{(\psi)} \triangleq \sum_{l \in \mathcal{N}_k \setminus \{k\}} a_{2,lk}^2 \Lambda_{l,i} R_{v,lk}^{(\psi)} \quad (26)$$

*C. Entry Selection Method*

In order to select $L$ out of $M$ entries of the intermediate estimates of each node at each iteration, the methods we utilized are comparable to the selection processes in stochastic and sequential *partial-update* schemes [9, 20, 21]. In other word, we use the same schemes as that introduced in [14]. Here, we just review these methods namely *sequential* and *stochastic* partial-diffusion.

In sequential partial-diffusion the entry selection matrices, $\Lambda_{k,i}$, is a diagonal matrix:

$$\Lambda_{k,i} = \begin{bmatrix} r_{1,i} & \cdots & 0 \\ \vdots & \ddots & \vdots \\ 0 & \cdots & r_{M,i} \end{bmatrix}, \quad r_{\ell,i} = \begin{cases} 1 & if \ \ell \in \mathcal{I}_{(i \, mod \, \bar{B})+1} \\ 0 & otherwise \end{cases} \quad (27)$$

with $\bar{B} = \lceil M/L \rceil$. The number of selection entries at each iteration is limited by $L$. The coefficient subsets $\mathcal{I}_i$ are not unique as long as they meet the following requirements [9]:

1. Cardinality of $\mathcal{I}_i$ is between 1 and $L$;
2. $\bigcup_{r=1}^{\bar{B}} \mathcal{I}_r = \mathcal{S}$ where $\mathcal{S} = \{1,2,...,M\}$;
3. $\mathcal{I}_r \cap \mathcal{I}_p = \emptyset, \forall r, p \in \{1,...,\bar{B}\}$ and $r \neq p$.

The description of the entry selection matrices, $\Lambda_{k,i}$, in stochastic partial-diffusion is similar to that of sequential one. The only difference is as follows. At a given iteration, $i$, sequential case one of the set $\mathcal{I}_r, r = 1, ..., \bar{B}$ is chosen in a predetermined fashion, whereas for stochastic case, one of the sets $\mathcal{I}_r$ is sampled at random from $\{\mathcal{I}_1, \mathcal{I}_2, ..., \mathcal{I}_{\bar{B}}\}$. One might ask why these methods are considered to organize these selection matrices. To answer this question, it is worth mentioning that the nodes need to recognize which entries of their neighbors' intermediate estimates have been propagated at each iteration. These schemes bypass the need for addressing (position in the vector) [5].

III. STEADY-STATE PERFORMANCE ANALYSIS

We now move on to examine the behavior of the general PDLMS implementations (19)-(21), and the influence of the mentioned perturbations on its convergence and steady-state performance. For this reason, we shall study the convergence of the weight estimates both in the mean and mean-square senses. In order to make the analysis tractable, we introduce the following assumptions on statistical properties of the measurement data and noise signals.

*Assumptions:*
1. The regression data $u_{k,i}$ are temporally white and spatially independent random variables with zero mean and covariance matrix $R_{u,k} \triangleq \mathrm{E}u_{k,i}^*u_{k,i} \geq 0$.
2. The noise signal $v_k(i)$, $v_{k,i-1}^{(w)}$ and $v_{k,i}^{(\psi)}$ are temporally white and spatially independent random variable with zero mean and covariance $\sigma_{v,k}^2$, $R_{v,k}^{(w)}$ and $R_{v,k}^{(\psi)}$, respectively. In addition, the quantities $\{R_{v,lk}^{(w)}, R_{v,lk}^{(\psi)}\}$ are all zero if $l \notin N_k$ or when $l = k$.

3. The regression data $\{u_{m,i_1}\}$, the model noise signals $v_n(i_2)$, and the link noise signals $v^{(w)}_{l_1 k_1, j_1}$ and $v^{(\psi)}_{l_2 k_2, j_2}$ are mutually independent random variables for all indexes $\{i_1, i_2, j_1, j_2, k_1, k_2, l_1, l_2, m, n\}$.
4. The step-sizes, $\mu_k \; \forall k$, are small enough such that their squared values are negligible.

We are interested in examining the evolution of the weight-error vectors. To do so, we let:

$$\widetilde{w}_{k,i} \triangleq w^o - w_{k,i} \quad (28)$$

We import the information from across the network into block vectors and matrices as follows:

$$\mathcal{R}_{u,i} \triangleq diag\{u^*_{1,i} u_{1,i}, u^*_{2,i} u_{2,i}, \dots, u^*_{N,i} u_{N,i}\} \quad (29)$$

$$s_i \triangleq diag\{u^*_{1,i} v_1(i), u^*_{2,i} v_2(i), \dots, u^*_{N,i} v_N(i)\} \quad (30)$$

$$v^{(w)}_i \triangleq col\{v^{(w)}_{1,i}, v^{(w)}_{2,i}, \dots, v^{(w)}_{N,i}\} \quad (31)$$

$$v^{(\psi)}_i \triangleq col\{v^{(\psi)}_{1,i}, v^{(\psi)}_{2,i}, \dots, v^{(\psi)}_{N,i}\} \quad (32)$$

$$\mathcal{M} \triangleq diag\{\mu_1 I_M, \dots, \mu_N I_M\} \quad (33)$$

$$\widetilde{w}_i \triangleq col\{\widetilde{w}_{1,i}, \dots, \widetilde{w}_{N,i}\} \quad (34)$$

Subsequently, some algebra demonstrates that

$$\boxed{\begin{aligned}\widetilde{w}_i = \mathcal{A}_{2,i}(I_{NM} - \mathcal{M}\mathcal{R}_{u,i})\mathcal{A}_{1,i-1}\widetilde{w}_{i-1} \\ -\mathcal{A}_{2,i}(I_{NM} - \mathcal{M}\mathcal{R}_{u,i})v^{(w)}_{i-1} - \mathcal{A}_{2,i}\mathcal{M}s_i - v^{(\psi)}_i\end{aligned}} \quad (35)$$

where

$$\mathcal{A}_{r,i} = \begin{bmatrix} A_{1,1,i} & \cdots & A_{1,N,i} \\ \vdots & \ddots & \vdots \\ A_{N,1,i} & \cdots & A_{N,N,i} \end{bmatrix}, \forall r \in \{1,2\} \quad (36)$$

$$A_{p,q,i} = \begin{cases} I_M - \sum_{l \in N_p \setminus \{p\}} a_{r,lp} \Lambda_{l,i} & \text{if } q = p \\ a_{r,qp} \Lambda_{q,i} & \text{if } q \in N_p \setminus \{p\} \\ O_M & \text{other wise} \end{cases} \quad (37)$$

*A. Mean Performance*

Taking expectation of both sides of (35) under *Remark* and *Assumptions*, we find that the mean error vector evolves according to the following recursion:

$$\mathbb{E}\widetilde{w}_i = Q_2 \mathbb{E}(I_{NM} - \mathcal{M}\mathcal{R}_u) Q_1 \mathbb{E}\widetilde{w}_{i-1} \quad (38)$$

where

$$Q_1 = \mathbb{E}[\mathcal{A}_{1,i}], \quad Q_2 = \mathbb{E}[\mathcal{A}_{2,i-1}] \quad (39)$$

From (38), we observe that in order for the recursion to be stable in the mean sense, the matrix $Q_2 \mathbb{E}(I_{NM} - \mathcal{M}\mathcal{R}_u) Q_1$ should be stable. Picking $Q_1$ and $Q_2$ which all their entries are real non-negative and all their rows add up to unity. Therefore, in the light of lemma 1 of [21], mean stability and asymptotic unbiasedness of the algorithm is guaranteed if the matrix $I_{NM} - \mathcal{M}\mathcal{R}_u$ is stable or equivalently if

$$|\lambda_{max}\{I_{NM} - \mathcal{M}\mathcal{R}_u\}| < 1 \quad (40)$$

where $\lambda_{max}\{.\}$ refers to the largest eigenvalue of a matrix. The set of the eigenvalue of $I_{NM} - \mathcal{M}\mathcal{R}_u$ is the union of the set of the eigenvalue of $I_M - \mu_k R_{u,k} \; \forall k$. Thus, (40) is satisfied when $|\lambda_{max}\{I_M - \mu_k R_{u,k}\}| < 1, \forall k$ or $|1 - \mu_k \lambda_{max}\{R_{u,k}\}| < 1, \; \forall k$. These inequalities determine the stability bounds for step-sizes as

$$0 < \mu_k < \frac{2}{\lambda_{max}\{R_{u,k}\}} \quad \forall k \quad (41)$$

*B. Mean-Square Performance*

The weighted variance relation for the error vector $\widetilde{w}_i$ can be obtained from the error recursion (35) as:

$$\boxed{\begin{aligned}\mathbb{E}\|\widetilde{w}_i\|^2_\Sigma = \mathbb{E}\|\widetilde{w}_{i-1}\|^2_{\Sigma'} + \mathbb{E}(s^*_i \mathcal{M} \mathcal{A}^T_{2,i} \Sigma \mathcal{A}_{2,i} \mathcal{M} s_i) \\ +\mathbb{E}(v^{*(w)}_{i-1} \mathcal{H}^*_i \Sigma \mathcal{H}_i v^{(w)}_{i-1}) + \mathbb{E}\left(v^{*(\psi)}_i \Sigma v^{(\psi)}_i\right)\end{aligned}} \quad (42)$$

where $\Sigma$ is an arbitrary positive semi-definite Hermitian matrix of size $NM \times NM$ and

$$\mathcal{H}_i \triangleq \mathcal{A}_{2,i}(I_{NM} - \mathcal{M}\mathcal{R}_{u,i}) \quad (43)$$

Furthermore, the matrix $\Sigma'$ can be expressed as

$$\Sigma' \triangleq \mathbb{E}\mathcal{B}^*_i \Sigma \mathcal{B}_i \quad (44)$$

with

$$\mathcal{B}_i \triangleq \mathcal{A}_{2,i}(I_{NM} - \mathcal{M}\mathcal{R}_{u,i})\mathcal{A}_{1,i-1} \quad (45)$$

The variance relation becomes

$$\begin{aligned}\mathbb{E}\|\widetilde{w}_i\|^2_\Sigma = \mathbb{E}\|\widetilde{w}_{i-1}\|^2_{\Sigma'} \\ + \Big(vec^T\{\mathcal{G}\}\mathcal{D}_2 + vec^T(\mathcal{H} R^{(w)}_v \mathcal{H}^*)\mathcal{D}_2 \\ + vec^T\left(R^{(\psi)}_v\right)\Big)vec(\Sigma)\end{aligned} \quad (46)$$

where

$$\mathcal{D}_1 = \mathbb{E}(\mathcal{A}^T_{1,i-1} \otimes \mathcal{A}^T_{1,i-1}) \quad (47)$$

$$\mathcal{D}_2 = \mathbb{E}(\mathcal{A}^T_{2,i} \otimes \mathcal{A}^T_{2,i}) \quad (48)$$

$$\mathcal{G} = \mathcal{M}\mathbb{E}[\boldsymbol{s}_i\boldsymbol{s}_i^*]\mathcal{M} = diag\{\mu_1\sigma_{v,1}^2 R_{u,1}, \ldots, \mu_N\sigma_{v,N}^2 R_{u,N}\} \quad (49)$$

$$\mathcal{H} \triangleq \mathbb{E}[I_{NM} - \mathcal{M}\boldsymbol{\mathcal{R}}_{u,i}] = I_{NM} - \mathcal{M}\mathcal{R}_u \quad (50)$$

$$R_v^{(w)} \triangleq \mathbb{E}\boldsymbol{v}_{i-1}^{(w)}\boldsymbol{v}_{i-1}^{*(w)} = diag\{R_{v,1}^{(w)}, \ldots, R_{v,N}^{(w)}\} \quad (51)$$

$$R_v^{(\psi)} \triangleq \mathbb{E}\boldsymbol{v}_i^{(\psi)}\boldsymbol{v}_i^{*(\psi)} = diag\{R_{v,1}^{(\psi)}, \ldots, R_{v,N}^{(\psi)}\} \quad (52)$$

Therefore, the steady-state weighted variance relation (46) becomes

$$\lim_{i\to\infty}\mathbb{E}\|\widetilde{\boldsymbol{w}}_i\|^2_{unvec\left[\left(I_{N^2M^2}-\mathcal{D}_1[\mathbb{E}(I_{NM}-\boldsymbol{\mathcal{R}}_{u,i}\mathcal{M})\otimes(I_{NM}-\boldsymbol{\mathcal{R}}_{u,i}\mathcal{M})]\mathcal{D}_2\right)vec(\Sigma)\right]}$$

$$= \left(vec^T\{\mathcal{G}\}\mathcal{D}_2 + vec^T\left(\mathcal{H}R_v^{(w)}\mathcal{H}^*\right)\mathcal{D}_2 + vec^T\left(R_v^{(\psi)}\right)\right)vec(\Sigma) \quad (53)$$

It is known that a recursion of type (53) is stable and convergent if the matrix $\mathcal{D}_1[\mathbb{E}(I_{NM} - \boldsymbol{\mathcal{R}}_{u,i}\mathcal{M}) \otimes (I_{NM} - \boldsymbol{\mathcal{R}}_{u,i}\mathcal{M})]\mathcal{D}_2$ is stable. All the entries of $\mathcal{D}_1$ and $\mathcal{D}_2$ are real non-negative and all its columns sum up to one. Moreover, the eigenvalue of $\mathcal{T} \otimes \mathcal{T}$ are square of the eigenvalue of $\mathcal{T}$. Therefore, stability of this matrix has the same conditions as the stability of $I_{NM} - \mathcal{R}_u\mathcal{M}$. This means that choosing the step-sizes is accordance with (41) makes the algorithm stable likewise in the mean-square sense hence convergent to steady state. The network MSD is defined as:

$$\text{MSD}^{network} \triangleq \lim_{i\to\infty}\frac{1}{N}\sum_{k=1}^{N}\mathbb{E}\|\widetilde{\boldsymbol{w}}_{k,i}\|^2 \quad (54)$$

Since we are free to choose $\Sigma$, we select it as $I_{N^2M^2} - \mathcal{D}_1[\mathbb{E}(I_{NM} - \boldsymbol{\mathcal{R}}_{u,i}\mathcal{M}) \otimes (I_{NM} - \boldsymbol{\mathcal{R}}_{u,i}\mathcal{M})]\mathcal{D}_2 = vec(I_{NM}/N)$ then the expression (53) gives

$$\boxed{\text{MSD}^{network}_{imperfect} = \frac{1}{N}\left[vec(\mathcal{G})\mathcal{D}_2 + vec(\mathcal{H}R_v^w\mathcal{H}^*)\mathcal{D}_2 + vec\left(R_v^{(\psi)}\right)\right]^T(I_{N^2M^2} - \mathcal{F})^{-1}vec(I_{NM})} \quad (55)$$

## IV. NUMERICAL STUDIES

### A. Simulation

In order to illustrate the PLDMS strategies performance under noisy information exchange, we consider an adaptive network with a random topology and $N = 10$ where each node is, in average, connected to two other nodes. The unknown parameter $w^o$ of length $M = 8$ is randomly generated. We adopt a uniform step-size, $\mu_k = 0.01$, The measurements were generated according to model (1), and regressors, $\boldsymbol{u}_{k,i}$, were chosen Gaussian i.i.d with randomly generated different diagonal covariance matrices, $R_{u,k}$. The additive noises at nodes are zero mean Gaussian with variances $\sigma_{v,k}^2$ and independent of the regression data. The traces of the covariance matrix regressors and the noise variances at all nodes, $Tr(R_{u,k})$ and $\sigma_{v,k}^2$, are shown in Fig. 2.

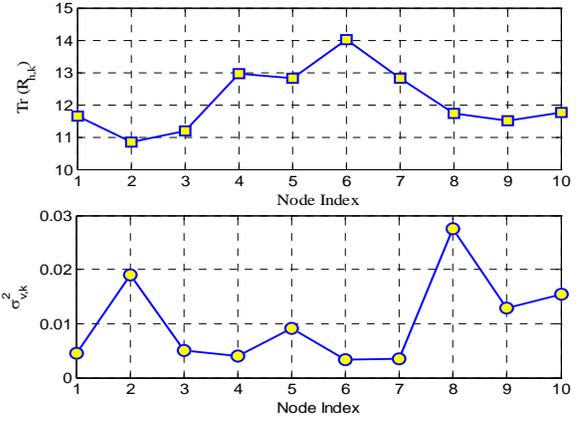

Figure 2. Covariance matrix trace of the input signal and the variance of the noise at each node.

We also use white Gaussian link noise signals such that $R_{v,lk}^{(w)} = \sigma_{w,lk}^2 I_M$ and $R_{v,lk}^{(\psi)} = \sigma_{\psi,lk}^2 I_M$. All link noise variances $\{\sigma_{w,lk}^2, \sigma_{\psi,lk}^2\}$ are randomly generated. The average power of each type of link noise across the network is 35 dB less than that of the model noise. In Fig. 3, we plot the experimental network MSD curves for ATC case ($A_1 = I_N$) of PDLMS algorithm using both sequential and stochastic partial diffusion schemes under noisy information exchange for different numbers of entries at each iteration, $M$. We use uniform weights for $\{a_{1,lk}, a_{2,lk}\}$ at combination phase at this stage. In Fig. 4, we compare network MSD learning curve of PDLMS

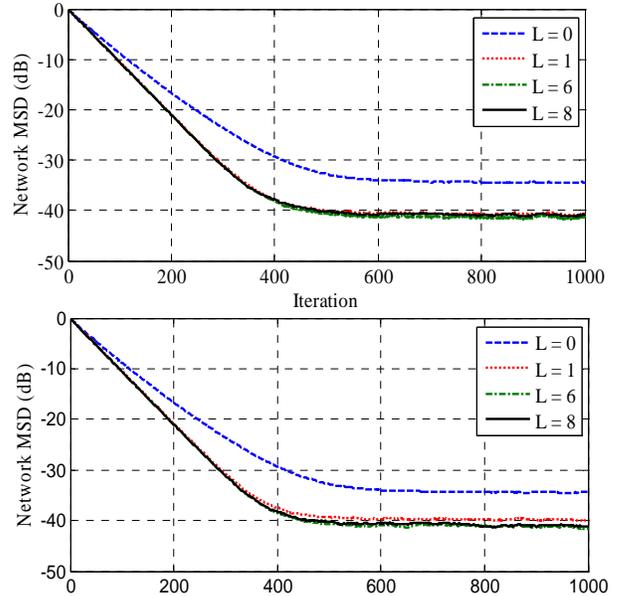

for both ATC and CTA ($A_2 = I_N$) under ideal and noisy links.

Figure 3. Network MSD learning curve of PDLMS for ATC under nosiy links (top) sequential (bottom) stochastic.

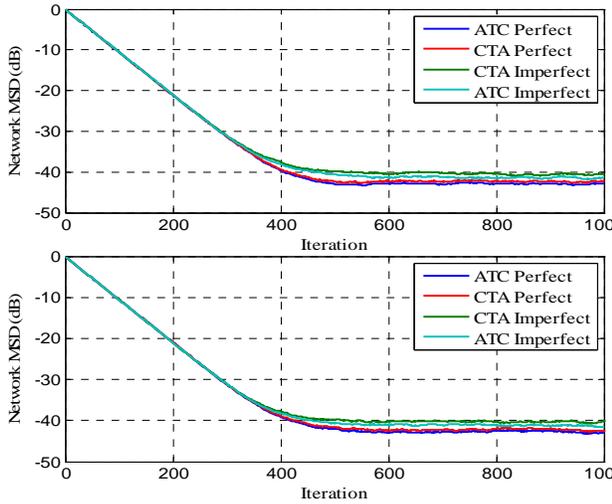

Figure 4. Network MSD learning curve of PDLMS for both ATC and CTA under ideal and noisy links (a) sequential (b) stochastic.

*B. Discussion*

From the simulation results, we make the following observations:

--First, in [5], authors emphasized that the PDLMS algorithm delivers a tradeoff between communications cost and estimation performance. However, Based on simulation results considering the assumption of noisy links add a new complexity to the optimization problem and the trade-off between communication cost and estimation performance in comparison to ideal links become unbalanced. Because, the more entries are broadcast at each iteration, the more perturbed weight estimates are interred in consultant phase.

--Second, the sequential partial-diffusion schemes outperform the stochastic partial-diffusion.

--Third, adaptive ATC strategy outperforms the adaptive CTA strategy for both perfect and imperfect cases.

V. CONCLUSION

In this work, we presented a general form of PDLMS algorithms, formulated the ATC and CTA versions of PDLMS under noisy Links, and investigated the performance of partial-diffusion algorithms under several sources of noise during information exchange for both sequential and stochastic schemes. We also illustrated that the PDLMS strategy could still stabilize the mean and mean-square convergence of the network with noisy information exchange. We derived analytical expressions for network learning curve MSD. The important result is that the noisy links are the main factor in performance degradation of a diffusion LMS strategy running in a network with imperfect communication. Furthermore, there is no direct relation between the MSD performance and number of selected entries under imperfect information exchange.